\documentclass[onecolumn, amsmath, amssymb, 12pt, nofootinbib]{revtex4-2}

\usepackage{graphicx}
\usepackage{dcolumn}
\usepackage{bm}
\usepackage{natbib}
\usepackage{breakurl}
\usepackage{xurl}

\begin{document}



\title{Modelling reliability under deep decarbonisation of the European electricity grid}

\author{J. Dunsmore$^{\textbf{*},}$$^{1,}$$^{3}$, L.M. Arthur$^{2,}$$^{4}$, R.S. Kemp$^{1,}$$^{4}$}

\address{\textbf{*}Corresponding Author: \textbf{jduns@mit.edu}}

\address{$^{1}$Department of Nuclear Science and Engineering,  MIT, Cambridge, MA, USA}

\address{$^{2}$Department of Earth, Atmospheric and Planetary Sciences,  MIT, Cambridge, MA, USA}

\address{$^{3}$MIT Plasma Science and Fusion Center, Cambridge, MA, USA}

\address{$^{4}$Laboratory for Nuclear Security and Policy, MIT, Cambridge, MA, USA}

\makeatletter
\renewcommand\frontmatter@abstractfont{\normalsize}
\makeatother

\begin{abstract}

Conventional feasibility studies of deep decarbonisation are often limited in their temporal scope, and are thus unable to draw conclusions about grid reliability over multi-decadal time periods. To address this problem, we introduce RESCORE, a fast and transparent model that uses 43 years of hourly weather data to evaluate both the cost and reliability characteristics of low-carbon electricity mixes. Applying RESCORE to the European electricity grid, we show that infrequent but extreme weather events play an outsized role in setting the cost for generation mixes without dispatchable backup: reliably meeting the last 1\% of demand accounts for about 36\% of the entire system cost when the mix contains only solar, onshore wind and storage. We also show that by including small amounts of low-cost, dispatchable generation (e.g., natural gas, hydro, or similar) the cost of a reliable, high-renewables grid can be drastically reduced.

\end{abstract}

\maketitle

\section{Introduction}
\label{sec1}

Europe has some of the most aggressive CO$_{2}$-reduction targets in the world, with the EU pursuing net zero carbon emissions by 2050 across its 27 member states ~\citep{EU_netzero} and several individual countries setting legally binding targets to reach the same milestone~\citep{UK_2019_netzero, European_countries_netzero, Uk_climate_change_act}. Due to their low and falling costs, wind and solar power are expected to play a major role in the decarbonisation of the electricity grid in Europe~\cite{IRENA_costs_NEW_2023, lazard_lcoe, EU_net_zero_energy}.

However, the inherent variability of wind and solar generation has led to questions about their potential to achieve very deep decarbonisation (near 100\%) while ensuring reliable grid operation~\cite{shaner, MIT_genx, shaner_long_duration_energy_storage, jenkins_deep_decarbonisation_difficulties, davis_netzero_review}. These concerns have motivated several studies on the feasibility of a 100\% renewable power sector \cite{brouwer2016_EU_power_sector_model, bussarEU_power_sector_modelling, bussarEU_power_sector_modelling_sensitivity, gils2017_EU_power_sector_model, Plexos_model_2019} and energy sector \cite{EU_net_zero_energy, brown2018_europe_energy_model, simoes2017_europe_energy_model, LUT_model_2019} in Europe. Such studies generally estimate wind and solar generation potential across the European continent using climate reanalysis datasets, and feed this into a power system model to optimise for the minimum-cost system that meets demand given technical constraints on generator capacity, transmission capacity, generator ramp-time and other factors. However, these models are computationally expensive and, as a result, are often limited in their temporal scope. For example, two widely-cited studies exploring 100\% renewable scenarios in Europe employed the LUT energy system model \cite{LUT_model_2019} and the PLEXOS model energy system model \cite{Plexos_model_2019}. Both studies were conducted using only a single year of weather data: the LUT study used a typical weather year \cite{LUT_introduction_model}, while the PLEXOS study used a performance-limiting weather year but included just one week of data per month during the year.

This reduction in temporal scope is necessary when using complex power sector models like PLEXOS and LUT, but it means  that the results obtained are not robust over multi-decadal time scales. The need for high temporal resolution, long-duration weather data in energy system planning is already well-documented \cite{need_for_longterm_data1, need_for_longterm_data2, ruggles_2024}, and recent work by Ruggles et.~al.\ has found that almost 40~years of weather data are required to ensure a reliable grid in systems without dispatchable backup \cite{ruggles_2024}.

Computationally expensive models force other limitations as well. The LUT model assumes perfect foresight of supply and demand in its simulations \cite{LUT_introduction_model}, which yields overly optimistic projections surrounding the levels of storage and backup needed to supplement a highly renewable grid. In addition, detailed sensitivity analyses are often impossible with computationally expensive models. While previous studies have explored the sensitivity of the system cost to single parameters like the permitted carbon intensity of the grid or individual technology costs \cite{MIT_genx, Europe_grid_sensitivity_study}, scanning across multiple inputs simultaneously (e.g., investigating how varying the carbon intensity would affect the overall system cost under many different technology cost assumptions) becomes prohibitively expensive.

In order to overcome these challenges, we introduce here the RESCORE (Renewable Energy System Cost Optimisation and Reliability Evaluation) Model. Inspired by the work of Shaner et.~al.\ \cite{shaner} and Tong et.~al.\ \citep{tong_paper}, RESCORE uses long-duration weather data to obtain accurate reliability estimates for highly-renewable grids. The model treats the grid as a single node, averaging supply and demand across the entire region of interest under the assumption of lossless transmission. RESCORE then simulates the single-node grid with hourly resolution from 1980--2022, varying the installed capacities of solar, onshore wind, storage, and dispatchable backup generation, and assuming a spatially homogenous distribution of renewable resources.

In this paper, we use RESCORE to give the first cost estimate of a highly-renewable grid in Europe with a realistic reliability constraint over a multi-decadal time period. Modern targets for grid reliability in Europe are 99.97\% (corresponding to an average of 3 hours per year in which supply cannot meet demand) \cite{europe_reliability}, but since desired grid reliability is an input to the RESCORE model, we are also able to explore for the first time how varying this reliability target would affect both the cost and make-up of a highly-renewable European grid. RESCORE returns full reliability information including the severity and duration of all outages, enabling these metrics to be analysed in detail. The speed of the model also allows for a rich sensitivity analysis in which the interplay between several different constraints and inputs can be investigated. We investigate here the effect of simultaneously varying the target grid reliability, the carbon intensity, and the cost of electricity storage on the overall system cost. These constraints are highly interdependent, and must be studied together using a tool like RESCORE in order to fully capture their impact on the system cost.

\section{Methods I - Data}
\label{section:methods_data}
The RESCORE model relies on hourly time series data for solar generation and wind generation over a 43-year period (1980--2022). These time-series were generated using raw weather data from the MERRA-2 reanalysis dataset \cite{MERRA_2_rad, MERRA_2_slv}. Hourly electricity demand data for the year 2022 was collated from the ENTSO-E Transparency Platform \cite{demand_data}.

MERRA-2 contains wind speed and solar surface flux measurements from 1980 onwards, at a temporal resolution of one hour and a spatial resolution of 0.5° $\times$ 0.625°, which translates to about one grid point every 50~km at European latitudes. For context, a total of 208 MERRA-2 grid points are located within France, while the entirety of the analysis area contains 2044 grid points.\\

\subsection{Calculating wind \& solar generating potential}

An estimate for the wind and solar generating potential at every grid was obtained using data from the MERRA-2 Radiation Diagnostics \cite{MERRA_2_rad} and Single-Level Diagnostics \cite{MERRA_2_slv} datasets. For solar generation, we assumed non-tracking panels and a standard efficiency of 21\% (defined as the efficiency under 1000 $W/m^2$ direct illumination at a panel temperature of 25~°C) \cite{solar_efficiencies}. To calculate the generation, we followed a similar method to Pfenninger and Staffell \cite{pfenninger_solar}, and Sandia National Laboratory \cite{Sandia_pvlib}---and we accounted for the temperature-dependence of panel efficiency using Huld's model \cite{huld_model}. Full details of the calculation can be found in Appendix~\ref{appendix:wind_and_solar_calcs}.

A similar calculation was performed for the wind generating potential. We assumed all wind turbines to have a turbine rotor diameter of 110~m, a hub-height of 100~m, a maximum rated power of 4.1~MW and cut-in and cut-out speeds of 3~m/s and 25~m/s respectively. Hellman's power law equation was used to extrapolate wind speeds recorded at 10m and 50m in the MERRA-2 dataset to the assumed hub height of 100~m \citep{wind_profile}. Again, full details can be found in Appendix~\ref{appendix:wind_and_solar_calcs}.

This process yielded two hourly-resolution generation profiles at every MERRA-2 grid point: a `wind power per turbine' and a `solar power per unit area'. We averaged over all 2044 points in the grid to obtain a single solar generation profile and a single wind generation profile for the whole of Europe, and then scaled these profiles to set the installed capacity of wind and solar in the baseline case equal to peak electricity demand in Europe (486 GW).

\subsection{Estimating electricity demand\\}

The European Network of Transmission System Operators for Electricity (ENTSO-E), which operates the synchronous grid in Europe, requires its members to publish hourly electricity demand data every year, and makes this information publicly available \citep{demand_data}. However, not all countries connected to the synchronous European grid (CESA) are members of ENTSO-E, and the UK, Turkey and the North African states in particular do not publish hourly demand data on the ENTSO-E platform. Data for the UK are available elsewhere \cite{uk_demand_data}, but Turkey and the North African states were excluded from our analysis due to data being unavailable. Some countries, such as Ukraine and Albania, use the ENTSO-E platform but do not reliably publish data. We only included in our analysis the 35 countries that provided demand data for more than 8700 hours of the year (out of 8760) in 2022---see Appendix~\ref{section:countries}. The borders of these countries define the locations of the MERRA-2 grid points used to calculate solar and wind generating potential.

For the remaining hours with missing demand data, we estimated these values using linear interpolation. To model demand over the full 43-year period of the MERRA-2 dataset, we duplicated the 2022 demand data for each year 1980--2022, as demand data is not available over the full period covered by MERRA-2. For a detailed discussion of this point, and further analysis of the consequences of using a single year of demand data, see the Supplementary Information.\\

\subsection{Storage \& Backup\\}

In order to analyse the effects of adding storage and dispatchable backup to a highly renewable electricity grid, we had to choose technologies to capture these characteristics. For storage, we chose lithium-ion batteries, since this is currently the lowest cost battery technology and has been widely studied in the context of future grid-scale electricity storage \cite{lithium_ion_grid_storage_1, lithium_ion_grid_storage_2}. We followed the cost estimates in \cite{storage_costs_NREL} for a 4-hour Li-ion storage system, and assumed that maintenance covered under the fixed annual O\&M costs ensures no degradation of the battery over its life cycle \cite{storage_costs_NREL}. For simplicity, we have assumed perfect round-trip efficiency and no self-discharge for the batteries. We would not expect any significant changes to our results if these terms were included since, in our threshold-based model (see Section \ref{sec:running_the_simulation}), storage inefficiencies would lead to dispatchable backup being deployed slightly earlier. This would increase fossil fuel emissions for each scenario slightly, but would likely not affect the system cost or the reliability of the grid.

We chose natural gas as the dispatchable resource in this analysis. Ideally, this dispatchable backup would be provided by a carbon-free source, but no current carbon-free electricity source has the low capital cost and flexibility required to be a suitable low-utilisation backup to a grid with high penetrations of wind and solar. Using natural gas also allows us to investigate the trade-offs between carbon intensity, system cost, and reliability in a future electricity grid.\\

\section{Methods II - The RESCORE Model}

The result of the procedure detailed in Section \ref{section:methods_data} is a dataset containing average solar generation, average wind generation and average electricity demand at hourly resolution over the 43-year period. We feed this dataset as an input to RESCORE, which uses a rules-based dispatch model to calculate the average grid reliability and the percentage of electricity produced by the dispatchable source for a given grid configuration. We then combine this with a cost model for system to find the minimum-cost blend given constraints on grid reliability and dispatchable generation deployment. While not essential to the model, including the constraint on dispatchable deployment  allows us to explore the cost and reliability trade-offs associated with different levels of decarbonisation if natural gas is used as the dispatchable generator. We do not include transmission in our cost estimates, since we do not model transmission, but achieving the perfect levels of interconnectivity assumed here would require major upgrades to the European transmission grid.

\subsection{Running the simulation}
\label{sec:running_the_simulation}

The simulation works as follows. At the start of each run, we specify five parameters:

\begin{enumerate}
    \item The renewables overbuild factor
    \item The relative proportion of wind and solar in the generation mix
    \item The total system storage capacity in GWh
    \item The total natural dispatchable generation capacity in GW
    \item A threshold to indicate when dispatchable generation should start being used (expressed as a percentage of the total system storage capacity)
\end{enumerate}

At each hour, we calculate solar and wind generation based on the overbuild factor (where a $\times$1 overbuild factor means the installed capacity for wind and solar equals peak demand) and the respective proportions of wind and solar in the generation mix. If residual load (demand minus renewable generation) is less than zero, then the excess renewable generation is used to replenish the storage up to its maximum value. If residual load is greater than zero and storage levels are greater than the threshold, then storage is called first to help meet demand. If residual load is greater than zero and storage levels are \textit{below} the threshold, then the dispatchable generators are turned on to meet demand and to restore the storage levels back to the threshold.

The simulation cycles through each hour in the 43-year period and outputs the reliability (the percentage of hours in which supply was sufficient to meet demand), as well as the percentage of total electricity that was generated by dispatchable generation.\\

Importantly, the threshold condition for dispatchable backup means that the model is completely myopic. The storage/generation capacity is exogenously specified at the start of each simulation, and at every time-step the system operator's direction is simply to: \\
\\
1) Never use dispatchable backup unless storage levels are below the threshold. \\
\\
2) Restore storage levels to the threshold as quickly as possible if they fall below the threshold. \\ \\

With a perfect foresight assumption, the operator would be able to `see' periods of low renewable generation far in advance, and would be able to deploy the dispatchable generation earlier to prevent blackouts. This would lead to unrealistically low system costs.

Furthermore, previous rules-based dispatch models for Europe \cite{europe_storage_and_balancing} and the United States \cite{US_storage_and_balancing} only deployed backup generation when storage levels fell to zero. However, it is plausible that deploying dispatchable backup \textit{before} the storage levels reach zero could improve grid reliability or reduce the backup capacity required. We include the dispatch threshold (input 5) as an optimisable parameter to allow for this possibility.

The simulation logic described here is displayed in the form of a flow diagram in the Supplementary Information.\\ \\ \\

\subsection{System cost calculations}
\label{sec:costcalc}
As well as running the grid simulation described in Section \ref{sec:running_the_simulation} for each set of generation capacity inputs, the corresponding annual system cost (neglecting transmission costs) is also calculated.

\begin{table}[t]
  \scriptsize
  \centering
  \begin{tabular}{|c|c|c|c|}
    \hline
      & Overnight Capital Cost (\$/kW) & O\&M Costs (\$/kW-year) & Lifetime (years)\\
    \hline
    Solar & 790 \cite{IEA_cost_figures} & 10 \cite{IRENA_costs} & 25 \cite{IEA_cost_figures}\\
    \hline
    Wind & 1540 \cite{IEA_cost_figures} & 40 \cite{IRENA_costs} & 25 \cite{IEA_cost_figures}\\
    \hline
    Storage & 200 (\$/kWh) \cite{storage_costs_NREL} & 10 (\$/kWh-year) \cite{storage_costs_NREL} & 15 \cite{storage_costs_NREL}\\
    \hline
    Dispatchable backup (natural gas) & 1000 \cite{IEA_cost_figures} & 20 \cite{EIA_costs} & 30 \cite{gas_stats}\\
    \hline
  \end{tabular}
  \caption{Cost estimates for each technology used in the analysis. Sources are IEA \cite{IEA_cost_figures}, IRENA \cite{IRENA_costs}, EIA \cite{EIA_costs} and NREL \cite{storage_costs_NREL, gas_stats}. Variable Operations \& Maintenance (O\&M) costs were assumed negligible for all technologies, and fuel costs for natural gas were neglected because its maximum utilisation was just 2\% of all electricity generation in the study. All cost data in $\$_{2025}$. Figure~\ref{fig: fig5} shows a sensitivity analysis to storage costs. The conversion to investment cost is described in the Methods section.\\}
  \label{tab:costdata}
\end{table}

To do this we take the overnight costs $I_{\text{ON}}$ from Table~1, and calculate the total investment cost per kW of capacity in the year the construction is assumed to start $P_0$. We account for the fact that construction happens over a finite period of time $T$, and assume a parabolic spending pattern

\begin{equation}
P_0 = I_{\text{ON}} \dfrac{6}{T^3}e^{-iT}\int_0^T \tau(T-\tau)e^{\pi \tau}e^{i(T-\tau)}\mathrm{d}\tau.
\label{eq:overnight_cap}
\end{equation}

The total construction time $T$ is assumed to be 2~years for all technologies, $\pi$ is the inflation rate (assumed to be 4\%) and $i$ is the nominal interest rate (assumed to be 8\%). 

Then the annual payment $A$ (capital recovery factor) per kW of capacity is

\begin{equation}
A = P_0 \left( \dfrac{e^{rL}{(e^{r}-1)}}{e^{rL} - 1} \right),
\label{eq:annual_repayment}
\end{equation}

where $r = i - \pi$ is the real interest rate. This loan repayment in \$/kW-year can be added to the fixed O\&M costs in \$/kW-year and multiplied by the installed capacity to give a total annual cost for each technology. Variable costs are assumed to be negligible, since wind and solar have no fuel costs and natural gas generates less than 2\% of all electricity for every scenario considered in this paper.

To calculate total annual system costs, the costs from the individual technologies (solar, wind, storage and natural gas) are summed together.

\subsection{Limitations of the RESCORE model}

As discussed in the Introduction, questions of long-term grid reliability and of complex trade-offs between reliability, cost and carbon-intensity necessitate the use of simplified models. RESCORE allows these questions to be addressed, but it has some limitations that will now be discussed. Because we do not consider transmission and only model four technologies, the total system costs reported here are not realistic estimates for true European system costs. Specifically, including transmission would increase costs while including alternative and existing sources of generation would decrease the overall system cost.

Furthermore, while the model uses 43 years of supply data, a lack of data availability means that we only include electricity demand data for the year 2022. The consequences of this are discussed in detail in the Supplementary Information, including an analysis of potential correlations between renewable-energy generation and electricity demand that are not captured by our analysis. We conclude that our results would not be meaningfully altered by the use of multiple years of weather data.

\begin{figure}[tb]
\centering
\includegraphics[width=0.7\columnwidth]{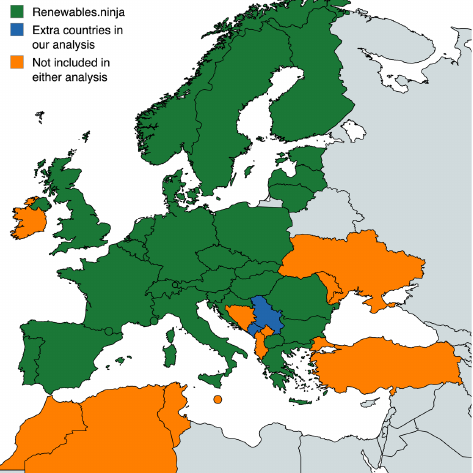}
\caption{Map of European countries included in the analysis. Countries shaded green are included in both our main analysis (presented in this paper) and in a validation analysis performed using the renewables.ninja dataset (see Appendix~\ref{appendix:validation} and Supplementary Information) \cite{staffell_wind, pfenninger_solar}. Countries shaded blue are included in our primary analysis but not in the renewables.ninja analysis. Countries shaded orange are connected to the Central European Synchronous Area (CESA) but are not included in either of our analyses.}
\label{fig: fig1}
\end{figure}

Finally, there is an extensive discussion in the literature surrounding the validity of reanalysis datasets like MERRA-2 for renewable generation studies \cite{staffell_wind, pfenninger_solar, MERRA_ERA_solar_biases, reanalysis_biases_wind}. Staffell and Pfenninger found discrepancies between wind and solar generation estimates from MERRA-2 and real-world generation data for European countries \cite{staffell_wind, pfenninger_solar}. To validate the conclusions of this paper, we have performed our analysis using both uncorrected MERRA-2 data as well as renormalised datasets for wind and solar published by Staffell and Pfenninger on the renewables.ninja platform \cite{staffell_wind, pfenninger_solar}. The countries covered in our original analysis and the countries covered in our validation analysis with the renewables.ninja dataset are shown in Figure~\ref{fig: fig1}. Full details of the differences between the datasets are included in Appendix~\ref{appendix:validation}, and a direct comparison of the results between the two datasets is included in the Supplementary Information. The results and conclusions presented in this paper were found to be robust to the choice of dataset used.

\section{Results}

\subsection{Required Overbuild \& Storage to Meet Reliability Targets}

\begin{figure}[tb]
\centering
\includegraphics[width=0.7\columnwidth]{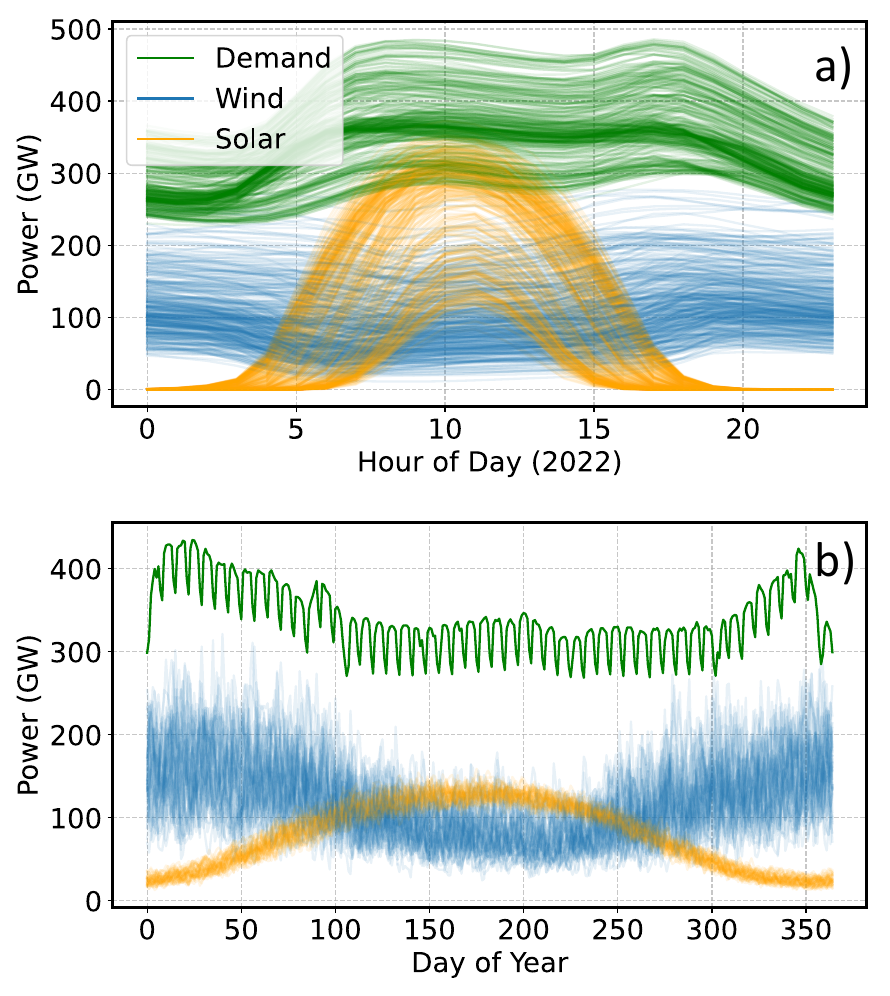}
\caption{a) Solar generation, wind generation and electricity demand for every day in 2022, with installed solar capacity and installed wind capacity both equal to peak demand (486 GW). Each line is a different day. b) Mean daily generation for every day of the year. Each line is a different year from 1980--2022. Electricity demand (green) is a single line because only one year of demand data was used. The average total electricity demand for the countries in our analysis is 340GW.}
\label{fig: fig2}
\end{figure}

\begin{figure*}[tb]
\centering
\includegraphics[width=\textwidth]{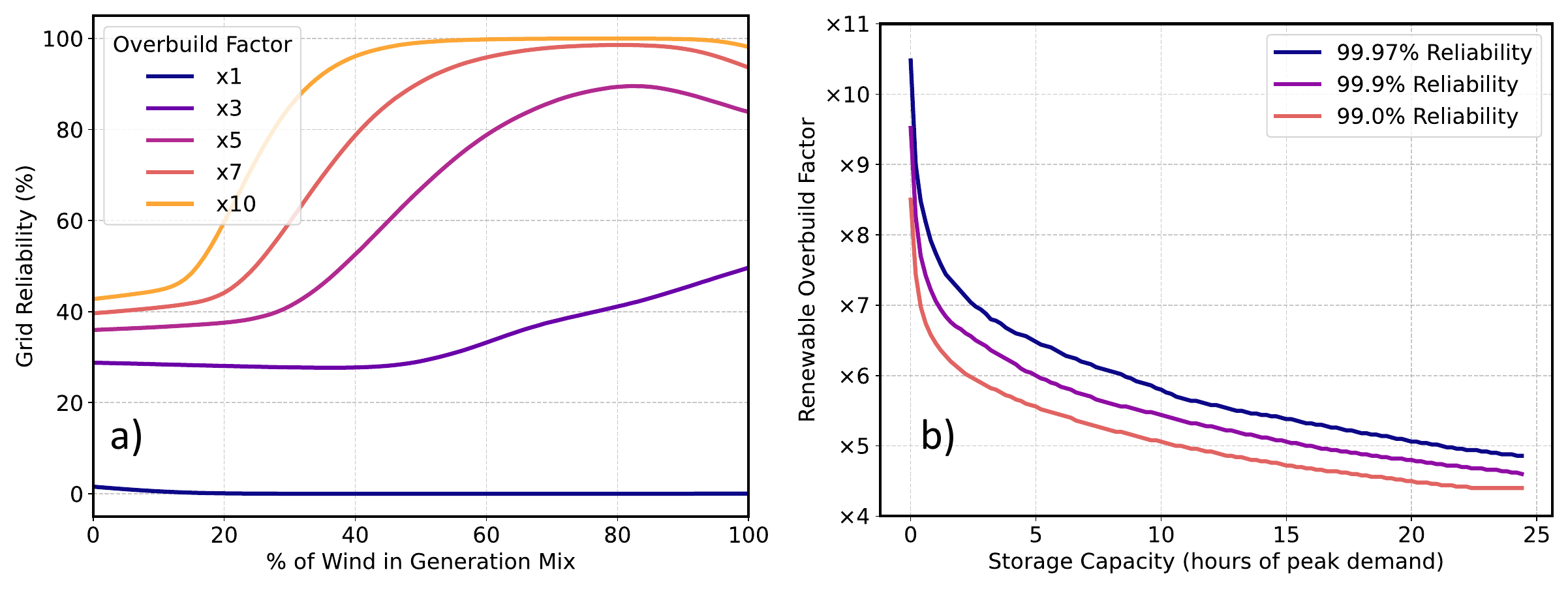}
\caption{a) Grid reliability levels as a function of the fraction of wind capacity in the generation mix (with no storage or dispatchable generation). An overbuild factor of $\times$1 means that the installed renewable capacity on the grid is equal to peak electricity demand (486 GW). b) Renewable overbuild factors required to reach three different grid reliability targets as a function of the storage capacity in the system.}
\label{fig: fig3}
\end{figure*}

We first investigate how the European-averaged wind and solar generation calculated from the MERRA-2 data vary with time. Figure~\ref{fig: fig2} shows the daily and seasonal variation in renewable generation over the 43-year period. Solar output is more predictable than wind output, with little year-to-year variation in the generation profile (Figure ~\ref{fig: fig2}b), but it experiences much larger diurnal and seasonal variation (summer generation is $\sim$x5 winter generation). Wind generation is remarkably stable over the course of a day, but has far less regularity in its seasonal profile than solar---thus making it harder to predict. Figure~\ref{fig: fig2}b also highlights the complementarity between wind and solar when averaged across the European continent, since solar is strongly peaked in summer and wind is peaked in winter. This supports previous studies, which have found complementarity between seasonal wind and solar generation in Europe \cite{prol2024_complementarities, Heide_2010_seasonal_mix}.

Figure~\ref{fig: fig2} also indicates that, without any overbuild, solar and wind rarely meet demand at any hour of the year. This is unsurprising, because current European capacity factors are between 10\%--20\% for solar and 10\%--30\% for onshore wind \cite{prol2024_complementarities}. In Figure~\ref{fig: fig3}a, we quantify this by plotting the obtained grid reliability over the 43-year period for different renewable overbuild factors as a function of the percentage of wind power in the generation mix. No storage or backup was included in the simulations for these cases. When the overbuild factor is $\times$1, meaning that the combined nameplate capacity of wind and solar equals peak demand (486 GW), demand is almost never met for any mix of wind and solar. Even when the installed wind and solar capacity is 4860 GW---ten times the peak electricity demand---the resulting grid reliability of 99.94\% still falls short of modern 99.97\% reliability targets. Diminishing returns are also observed at high overbuild factors: increasing the overbuild factor from $\times$7 to $\times$10 of peak demand only boosts grid reliability 1.4\% from 98.54\% to 99.94\%.\footnote{The interpretation of the renewable overbuild factors presented here is different from the interpretation in Shaner et.~al. \cite{shaner} and Tong et.~al. \citep{tong_paper}. In those papers, $\times$1 overbuild is defined as `average renewable generation = average demand', whereas in our analysis $\times$1 overbuild is defined as `installed renewable capacity = peak demand', which we believe is a more useful metric in helping to size the system.}

Adding some energy storage (but no dispatchable generation) into the mix rapidly reduces the overbuild factors needed to reach target levels of grid reliability. Figure~\ref{fig: fig3}b plots the renewable overbuild factor required to meet different reliability targets as a function of the storage capacity in the system. For a 99.97\% reliability target, 1458 GWh of storage (equivalent to 3 hours at peak demand) is sufficient to reduce the required overbuild factor from $\times$10.6 to $\times$7. However, once storage capacity extends beyond a few hours of peak demand, it has less of an effect on the required overbuild factor. A full 24 hours of storage (11,664 GWh) only reduces the required overbuild factor for 99.97\% reliability to $\times$5.

Installed battery storage capacity in Europe currently sits at about 17.2 GWh \cite{Solar_Power_Europe_Report}, which corresponds to just two minutes of peak demand. However, this number has doubled every year since 2020, and significant additional storage capacity is planned. Our results suggest that the first 1000 GWh of storage yield the greatest benefits to the system, with diminishing returns after this. Interestingly, pumped hydropower storage \textit{already} provides $\sim$2000 GWh in Europe---although 38\% of this is located in Norway alone and a further 17\% is in Switzerland \cite{european_pumped_hydro_IEA}. Geopolitical issues notwithstanding, the large-scale storage provided by pumped hydropower will be a crucial component in a future, highly renewable European electricity grid.

We also observe in the high ($>$99.9\%) reliability scenarios that reliability issues are generally caused by a small number of extreme events: in the 99.97\% reliable case with 24 hours of energy storage, only 3 years out of 43 see any outages at all. This highlights the need for multi-decadal analyses when considering the reliability of highly-renewable electricity grids.

\subsection{Cost-optimised scenarios}

Figure~\ref{fig: fig3} makes clear that some form of dispatchable generation will likely be needed to avoid prohibitively high renewable overbuild factors in a wind/solar dominated grid. In this section, we add dispatchable backup into the simulations in the form of natural gas peaking plants, and find the cost-optimised mix for solar, wind, gas and storage under different constraints on carbon intensity, grid reliability, and storage cost.

As described in the Methods section, we perform the cost-optimisation not just over generation and storage capacity, but also over the natural gas deployment threshold (see Section \ref{sec:running_the_simulation}). Essentially, this threshold parameter sets the balance between reliability and carbon-constraints. A higher threshold will reduce the risk of blackouts, while a lower threshold will reduce the amount of electricity that is generated by natural gas and thus reduce the carbon intensity of the grid.

The model is sufficiently cheap to run that we can scan over all five input parameters with high resolution. After this, each set of input parameters can be labelled with an associated cost, average reliability over the 43-year period and percentage of electricity generated by dispatchable backup. We then choose the optimum system configuration as the lowest-cost set of five inputs that satisfies given constraints on the reliability and backup generator utilisation.\\

\subsubsection{System costs vs.\ grid reliability}

\begin{figure*}[htbp]
\centering
\includegraphics[width=\textwidth]{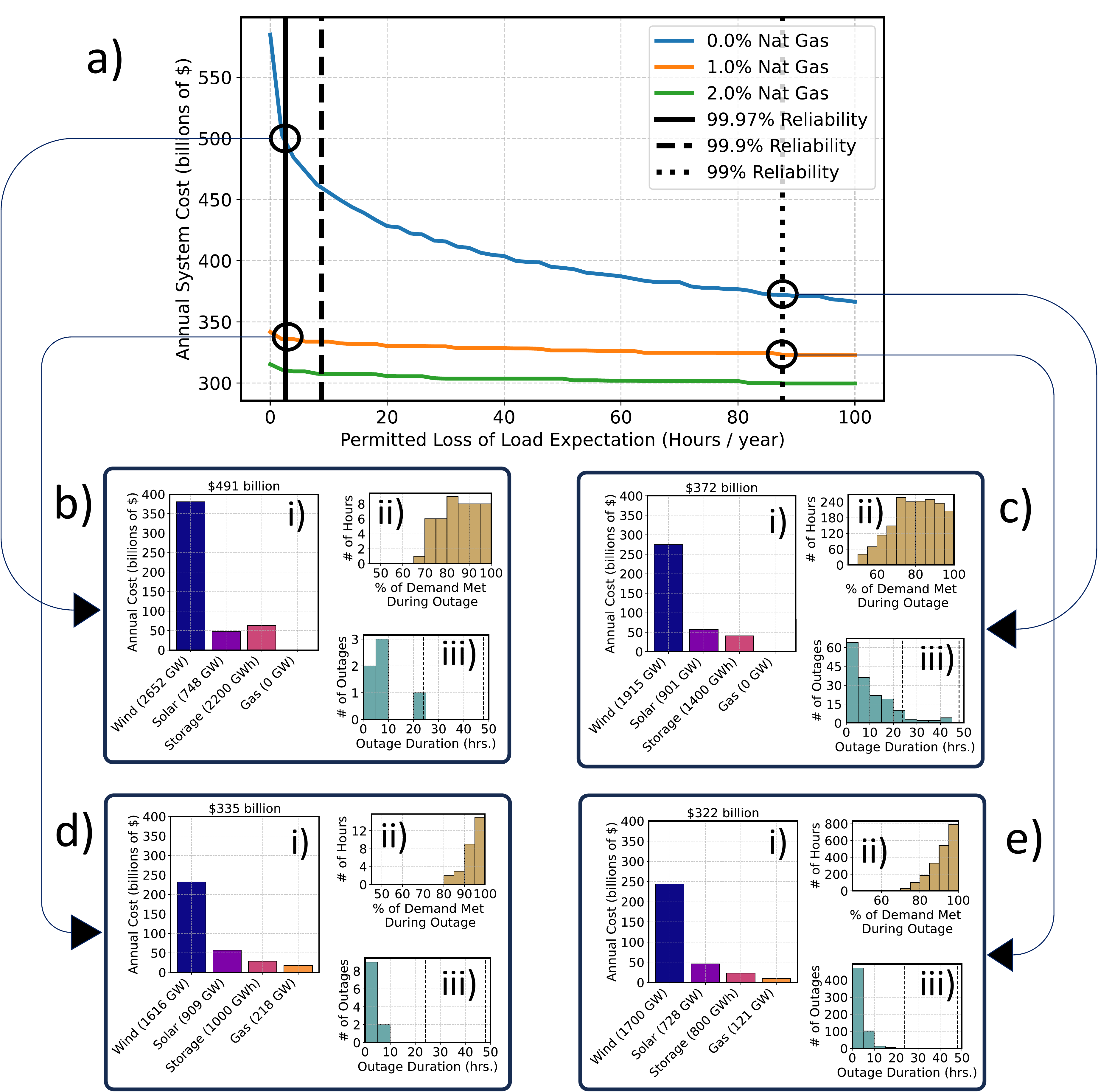}
\caption{a) Optimised annual system cost versus permitted Loss of Load Expectation per year. Loss of Load Expectation is defined as the average number of hours per year in which total demand exceeds total supply. Each subplot shows specific details about a given scenario: b) 0\% of generation from natural gas at 99.97\% grid reliability; c) 0\% from natural gas at 99\% grid reliability; d) 1\% from natural gas at 99.97\% grid reliability; e) 1\% from natural gas at 99\% grid reliability. For each scenario, i) shows the cost-breakdown of the generation mix, ii) shows the percentage of total demand met during each outage (a measure of the fraction of the load that can still be served through rolling blackouts or, inversely, a measure of the demand curtailment required to avoid the outage), and iii) shows the histogram of outage durations over the 43-year simulation period.}
\label{fig: fig4}
\end{figure*}

Figure~\ref{fig: fig4} shows how the annual system cost varies as a function of the reliability target for 3 different constraints on natural gas usage, with the different colours in Figure~\ref{fig: fig4}a representing different levels of permitted natural gas generation. Figures~\ref{fig: fig4}b--e contain further details about the cost-optimised results for the four circled cases in Figure~\ref{fig: fig4}a. 

The blue line in Figure~\ref{fig: fig4}a indicates that, in a European energy system containing only wind, solar and storage, the cost difference between a 99\% reliable grid and a 100\% reliable grid is about \$200 billion annually---or 36\% of the entire system cost. As soon as dispatchable backup is added into the mix, not only are major cost savings achieved, but the cost savings associated with relaxing the grid reliability target are also reduced. When just 1\% of total electricity generation is permitted to come from the dispatchable source, the optimal cost of a 99.97\% reliable grid falls from \$491 billion annually to \$335 billion annually. Furthermore, with 1\% of electricity from dispatchable backup, this 99.97\% reliable grid can be achieved at relatively little additional cost compared with a 99\% reliable grid (an additional \$13 billion per year).

These conclusions are consistent with the results in \cite{MIT_genx}, which found that the costs of decarbonising the power sector in the absence of firm resources grow rapidly for the last few percent of demand.

From the cost breakdowns in Figure~\ref{fig: fig4}di and Figure~\ref{fig: fig4}ei, it is clear that the \$13 billion cost of increasing the reliability from 99\% to 99.97\% is the result of a doubling of natural gas capacity (not usage)---from 121 GW to 218 GW---while the rest of the infrastructure required in the cost-optimised case remains largely the same. When dispatchable backup is allowed into the system, the installed backup capacity becomes the primary lever for increasing grid reliability. The same picture emerges when we examine the 2\% natural gas case in Figure~\ref{fig: fig4}a, with the \$10 billion difference in cost (\$299 billion annually to \$309 billion annually) between a 99\% reliable grid and a 99.97\% reliable grid being solely due to a doubling in the dispatchable backup capacity.

Another difference between Figure~\ref{fig: fig4}di and Figure~\ref{fig: fig4}ei is an increase in the optimum threshold level determined by the simulation---from 0\% to 5\%. Unsurprisingly, backup capacity must start being deployed \textit{before} the storage levels fall to zero in order to reach very high reliability levels. We would expect the optimum threshold level to rise further under stricter reliability targets or weaker carbon constraints.

It is also interesting to note that wind significantly dominates the overall cost in all scenarios, and that all scenarios rely on large renewable overbuild factors ($\times$7 for Figure~\ref{fig: fig4}b, $\times$5.8 for Figure~\ref{fig: fig4}c, $\times$5.2 for Figure~\ref{fig: fig4}d and $\times$5 for Figure~\ref{fig: fig4}e, with overbuild defined as nameplate capacity divided by peak demand), even when storage and dispatchable backup are permitted in the system. The high share of wind generation is likely due to a combination of two factors. The first is that European electricity demand is about 30\% higher in winter than in summer (as shown in Figure~\ref{fig: fig2}b), meaning that it aligns with the seasonal peak in wind generation rather than solar generation. The second reason is that wind generation experiences less variation than solar, with generation during the peak months approximately $\times$2 higher than during the trough (as opposed to approximately $\times$5 for solar). It makes sense, then, that a wind-dominated generation mix is most able to track the demand curve in Figure~\ref{fig: fig2}b with the minimum amount of overbuild. This optimum mix could change in the future if increased air conditioning use leads to a shift in peak demand from winter to summer.

To add context to the installed capacity numbers in Figure~\ref{fig: fig4}, total installed wind capacity in Europe was 272 GW at the end of 2023---with annual additions of 20 GW in 2022 and 18 GW in 2023 \cite{wind_europe_2023, wind_europe_2022}. At a build-rate of 20 GW per year, total installed capacity would be 812 GW in 2050, which is at least a factor of two lower than the wind capacity required to achieve a reliable grid in all our scenarios.

European installed solar capacity in Europe in 2023 was similar to wind (263 GW), but annual additions in 2023 were much higher than at 56 GW \cite{solar_europe_2023}. At this build-rate, there would easily be enough solar capacity to cover all of our scenarios in Figures~4 and 5. This highlights a fundamental tension in the deployment of renewable energy in Europe: solar projects are more profitable than wind at an individual level (hence the rapid roll-out), but the overall system costs are reduced when wind is prioritised.

Our results also show that the addition of some form of dispatchable backup into the mix affects the nature of the system outages. Comparing like-reliability cases (4b with 4d and 4c with 4e), it is clear that for cases in which 1\% of electricity is allowed to come from the dispatchable source, the typical duration of an outage is shorter and the percentage of demand met during each outage is higher. Indeed, it is possible that many of the outages in which greater than 90\% of demand is met could be effectively managed in the future with suitable demand-response measures.\\

\subsubsection{System costs vs.\ storage capital costs}

\begin{figure*}[tb]
\centering
\includegraphics[width=\textwidth]{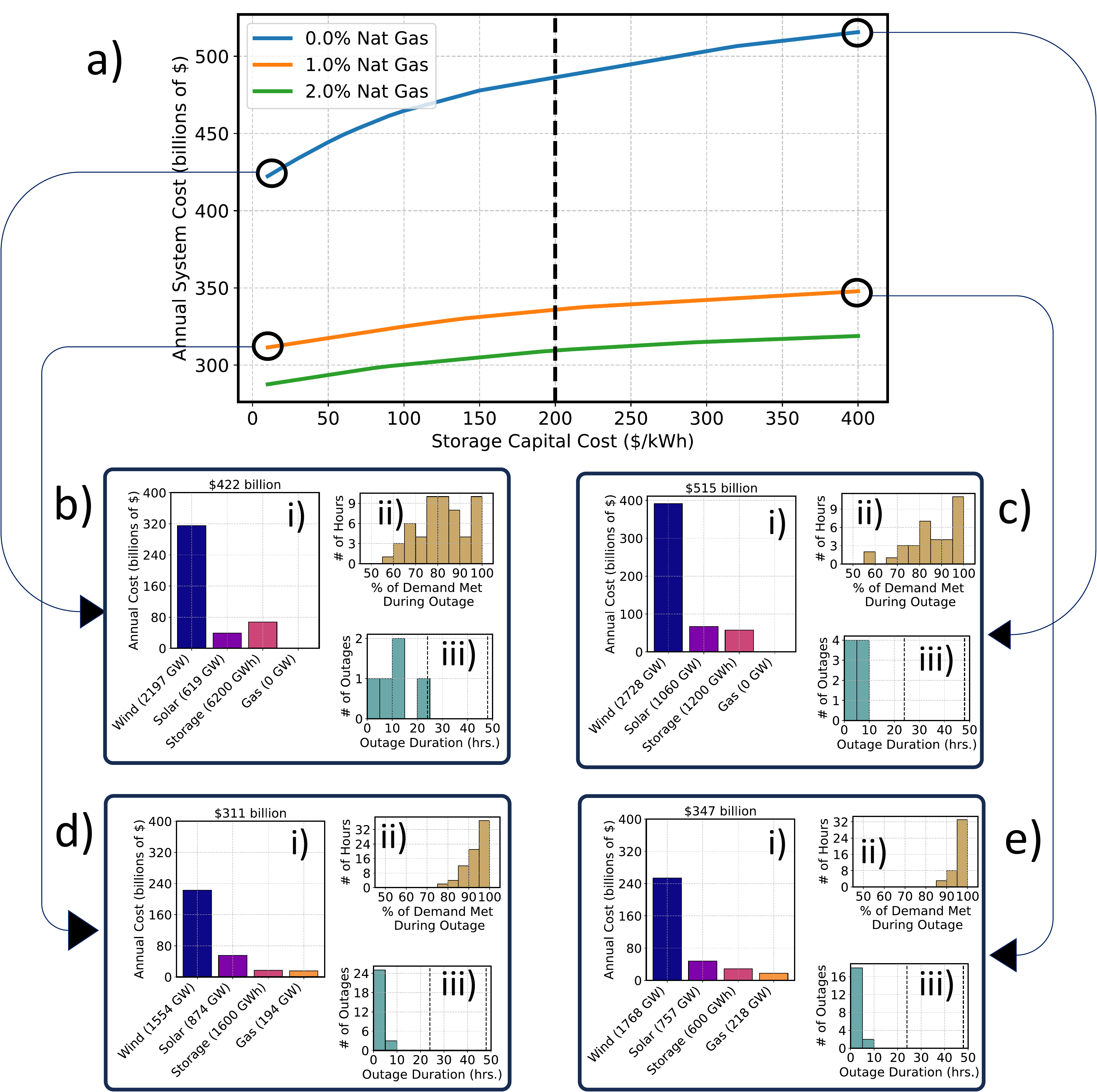}
\caption{a) Optimised annual system cost versus storage capital cost for a 99.97\% reliable grid. b) is 0\% natural gas at storage cost of \$10 per kWh; c) is 0\% natural gas at storage cost of \$400 per kWh; d) is 1\% natural gas at storage cost of \$10 per kWh; e) is 1\% natural gas at storage cost of \$400 per kWh. The subplots display the same information as in Figure~\ref{fig: fig4}}
\label{fig: fig5}
\end{figure*}

Figure~\ref{fig: fig5} shows instead how the optimised system cost for a 99.97\% reliable grid varies as a function of the storage capital cost. In the results shown in Figure~\ref{fig: fig4}, we assumed \$200 per kWh (following \cite{storage_costs_NREL}), but there is a large uncertainty in this estimate for 2050 battery storage costs. We explore here how the total system cost is affected by the value chosen for the storage capital cost. Again, we include details for four specific example cases in Figures~\ref{fig: fig5}b-e.

These results show that the cost reductions arising from the inclusion of a firm and flexible backup generator significantly outweigh the cost reductions from any decrease in storage cost. For a grid with no dispatchable backup, reducing the capital cost of storage from \$400 per kWh down to \$10 per kWh results in annual system cost savings of \$93 billion (\$515 billion to \$422 billion). If, instead, the capital cost of storage is kept fixed at \$400 per kWh but natural gas is allowed to generate 1\% of total electricity, then annual system cost savings of \$168 billion are achieved (\$515 billion to \$347 billion). Indeed the cost savings that result from a further 1\% increase in gas generation are comparable to the cost savings from reducing the storage capital costs from \$400 per kWh to \$10 /kWh again. A grid with 1\% of electricity from natural gas and storage at \$10 per kWh costs \$311 billion, while a grid with 2\% of electricity from natural gas but storage at \$400 per kWh costs just \$318 billion---only a \$7 billion annual increase in the total system cost despite a 40-fold increase in the storage capital cost.

Furthermore, just as Figure~\ref{fig: fig4} shows that adding dispatchable backup makes the system cost less sensitive to the target grid reliability, Figure~\ref{fig: fig5} shows that adding dispatchable backup makes the system cost less sensitive to the capital cost of storage. Without backup, increasing storage capital costs from \$10 per kWh to \$400 per kWh increases system costs by 22\%, compared to a 12\% increase when gas is allowed to generate 1\% of all electricity.

An important point to note is that non-negligible storage capacity (amounting to a couple of hours of peak demand) is still included in the cost-optimised scenario at \$400 per kWh, even when dispatchable backup is permitted. This suggests that some storage is crucial to a low-cost system and supports the conclusions of \citep{europe_storage_and_balancing}, which found that adding small amounts of storage to the grid dramatically reduces the energy that must be covered by dispatchable backup resources. If a direct, carbon-free substitute for gas can be obtained, then the role for storage to play will likely be very small. However, if natural gas remains the backup generator of choice, then storage will be crucial to help limit the use of natural gas to cases where it is absolutely necessary.

The patterns of outages under varying storage costs are similar to those seen in Figure~\ref{fig: fig4}. For cases 5d and 5e with dispatchable backup permitted, there are no outages over the entire 43-year period in which less than 70\% of demand is met. Without backup, however, cases 5b and 5c show that hours where less than 70\% of demand is met do occur. Furthermore, Figure~\ref{fig: fig5}biii shows one outage in which supply falls short of demand for over 24 hours continuously. When dispatchable backup is added to the system, no outage lasts longer than 10 hours.

Counter-intuitively, in cases with equal dispatchable backup constraints (5b/5c and 5d/5e), reducing the cost of storage causes the intensity of outages to worsen. The case shown in 5c (high storage cost) has both a higher proportion of demand met during outages \textit{and} fewer long-duration outages relative to 5b (low storage cost). This is because more wind and solar capacity is deployed in the high-storage-cost case to compensate for the reduced storage capacity. When storage levels fall to zero and an outage occurs, the greater wind/solar in the high-storage-cost case means that a greater proportion of demand can be met.

\section{Conclusions}

We have simulated a renewables-dominated European electricity grid using the RESCORE model: a new long-duration, high temporal resolution grid model, and have found that a reliable grid will require high levels of renewable overbuild---with installed peak generating capacity greater than four times the peak demand. Even so, we find that cost-minimised systems require both storage \textit{and} dispatchable backup generation to achieve reliability targets. A large quantity of one cannot compensate for the complete absence of the other without incurring extreme penalties in system costs.

While backup generation could be provided by a variety of technologies, natural gas is likely to be the preferred choice in the short-to-medium term due to its low capital cost. Our results show that natural gas backup can yield large ($>$30\%) reductions in the system cost even at extremely low utilisation rates (producing just 1--2\% of all electricity).

Our study suggests a number of important policy findings. First, grid optimisation studies that force a zero-emissions target are likely to produce conclusions in terms of technology mix, cost, and feasibility that deviate significantly from the true social optimum, as disallowing very small emissions has a major impact on cost and reliability. As such, finding ways to preserve some legacy generators as backup capacity may be valuable if a renewables-dominated grid is to be pursued.

Second, high-storage scenarios may not be the right approach to coping with renewable variability. Not only were high-storage scenarios expensive, they also resulted in longer outage durations and lower fractions of load being served during outage conditions (e.g., through rolling blackouts). In contrast, scenarios with lower storage capacity and thus greater reliance on excess and/or backup generation give significantly better outage performance. The corollary is that that a simple grid-reliability factor is not sufficient for optimising the technology mix. A more appropriate approach might be to consider the cost of unserved load, perhaps increasing that cost with outage duration.

Third, we find that wind substantially dominates all cost-optimised scenarios that we analysed, with the grid needing approximately twice the amount of wind capacity relative to solar capacity. This has important near-term implications for grid planning in Europe, given that solar generation is currently being installed at a much faster rate.

Finally, it is important to note that the cost estimates presented in this study are illustrative. While they provide insight into cost differentials across different technology scenarios, they do not represent total system costs. A comprehensive estimate would need to account for sunk infrastructure costs, transmission expansion, and a broader range of generation technologies.

\appendix

\section{Full details of Solar \& Wind generation calculations}
\label{appendix:wind_and_solar_calcs}

\subsection{Solar}

In order to calculate solar generation at every grid point, we used the incident short-wave radiation at ground level (variable name: SWGDN) to compute direct illumination and the incident radiation at the top of the atmosphere to compute diffusion illumination (variable name: SWTDN). Both variables are contained within the MERRA-2 Radiation Diagnostics dataset \cite{MERRA_2_rad}. We used the 2-metre ambient air temperature (variable name: T2M) from the Single-Level Diagnostics dataset \cite{MERRA_2_slv} to account for the temperature-dependence of cell efficiency. \footnote{Note that SWGDN from the Radiation Diagnostics dataset differs from the SWGDN values in the Land Surface Forcings dataset \cite{MERRA_2_lfo} as the incident radiation is averaged over the total area of each grid cell in the Radiation Diagnostics dataset, while it is only averaged over the land area in each grid cell in the Land Surface Forcings dataset.}

To convert these variables into a solar output, we assumed non-tracking solar panels and followed a similar method to Pfenninger and Staffell \cite{pfenninger_solar}, and Sandia National Laboratory \cite{Sandia_pvlib}. First, we calculated the in-plane solar irradiation of the panel

\begin{equation}
I_{tot} = I_{dir, in\text{-}plane} + I_{dif, in\text{-}plane} + I_{refl, in\text{-}plane},
\label{eq:in_plane_irrad}
\end{equation}

where the total in-plane irradiation $I_{tot}$ is simply the sum of direct, diffuse and ground-reflected contributions. The in-plane component of ground-reflected radiation can be estimated as

\begin{equation}
I_{refl, in\text{-}plane} = \textit{GHI} \times \frac{a(1-\cos{t})}{2},
\label{eq:reflected_radiation}
\end{equation}

where \textit{GHI} is the Global Horizontal Irradiance (equivalent to SWGDN from MERRA-2), $a$ is the albedo, assumed to be a constant 0.3, and $t$ is the panel-tilt angle. We assumed that all panels were placed with the optimal tilt of $t$ = 0.76 $\times$ lat + 3.1 \cite{pfenninger_solar}.

In order to calculate $I_{dif, in\text{-}plane}$ in Eq. \ref{eq:in_plane_irrad}, we first estimated the diffuse fraction of the total \textit{GHI} using the logistic model proposed in \cite{diffuse_radiation_equation}: 

\begin{equation}
\textit{DHI} = \frac{\textit{GHI}}{1 + e^{-5.0033 + 8.6025 k_t}}.
\label{eq:diffuse_irrad}
\end{equation}

Here, $k_t$ is the clearness index, which we estimated as the ratio of the ground-level radiation (SWGDN) to extraterrestrial radiation (SWTDN), and \textit{DHI} is the Diffuse Horizontal Irradiation---the diffuse irradiation measured normal to the Earth's surface. This was converted to an in-plane value via

\begin{equation}
I_{dif, in\text{-}plane} = \textit{DHI} \times \frac{1+\cos{(t)}}{2}.
\label{eq:diffuse_inplane}
\end{equation}

To obtain $I_{dir, in\text{-}plane}$ in Eq. \ref{eq:in_plane_irrad}, we calculated the direct component of the solar radiation using

\begin{equation}
\textit{DNI} = \frac{\textit{GHI} - \textit{DHI}}{\sin{(h)}},
\end{equation}

where $h$ is the solar altitude angle and \textit{DNI} is the Direct Normal Irradiance (the solar irradiance along the direct path of the Sun's rays).

The in-plane component of the direct radiation, $I_{dir, in\text{-}plane}$, is then given by

\begin{equation}
I_{dir, in\text{-}plane} = \textit{DNI} \cos{(\textit{AOI})},
\label{eq:direct_inplane}
\end{equation}

where the angle of incidence, AOI, is the angle between the Sun's rays and the normal vector to the panel. It is defined as
\begin{equation}
\cos{(\textit{AOI}}) = \sin{h}\cos{t} + \cos{h}\sin{t}\cos{(a_p - a_s)},
\label{eq:AOI_formula}
\end{equation}

where $a_p$ is the panel-azimuth angle and $a_s$ is the solar-azimuth angle. We assumed that all panels were south facing with $a_p = 180$°.

We accounted for the temperature dependence of solar panel efficiency following Huld \cite{huld_model}, assuming crystalline silicon solar cells and a relationship between panel temperature and ambient air temperature of $T_{panel} = T_{amb} + 0.035\times I_{in\text{-}plane}$, with ambient air temperature from the MERRA-2 T2M variable.

We assumed a standard efficiency (at $I_{in\text{-}plane}$ = 1000 $W/m^2$ and $T_{panel}$ = 25~°C) of 21\% \cite{solar_efficiencies} for all solar panels. The power output $P$ of the panel in $W/m^2$ can then be calculated as 

\begin{equation}
P = 0.21 \eta_{rel} I_{in\text{-}plane},
\label{eq:solar_output}
\end{equation}

where the relative efficiency $\eta_{rel}$ is the temperature-dependent effect from \cite{huld_model}.

\subsection{Wind}

For wind generating potential, we calculated the wind power that would be generated over the 43-year period by a typical modern wind turbine placed in each grid cell. We assumed a turbine rotor diameter of 110~m, a hub-height of 100~m, a maximum rated power of 4.1~MW and cut-in and cut-out speeds of 3~m/s and 25~m/s respectively.

The power output of a single wind turbine is given by

\begin{equation}
P = \dfrac{1}{2} \rho_{air} A C_p v^{3},
\label{eq:wind_power}
\end{equation}

where $\rho_{air}$ is the density of air, $A$ is the area swept out by the wind turbine blades, $v$ is the wind speed in metres per second and $C_p$ is the performance coefficient of the turbine---which we assume to be 40\% in this analysis \citep{wind_turbines_europe_data}.

To estimate $v$ at the hub height of 100~m, we used the 10-metre wind speeds (variable names V10M and U10M) and the 50-metre wind speeds (V50M and U50M) from the MERRA-2 Single-Level Diagnostics dataset \cite{MERRA_2_slv}. A power law relationship of the form 

\begin{equation}
\dfrac{v_{h_1}}{v_{h_2}} = (\dfrac{h_1}{h_2}) ^ a
\label{eq:wind_speed}
\end{equation}

was then used to extrapolate these wind speeds to the hub height of 100~m \citep{wind_profile}. In Eq.~\ref{eq:wind_speed}, $v_{h_1}$ and $v_{h_2}$ are the wind speeds at heights $h_1$ and $h_2$, and $a$ is a constant that depends on site-specific geography. We estimated the atmospheric density $\rho_{air}$ at ground level from the the 2-metre air temperature (variable name T2M) and the surface pressure (variable name PS), and extrapolated to the hub height of 100~m using the barometric equation \cite{barometric}. We combined Eq.~\ref{eq:wind_power} with the constraints on turbine operation (cut-in speed, cut-out speed, and maximum output) to create a piece-wise model for wind output at every grid square:

\begin{equation}
    P = \begin{cases} 
    0 & v\leq 3\\
    \min(\dfrac{1}{2} \rho_{air} A C_p v^{3}, 4.1) & 3<v<25 \\
    0 & v \geq 25.
    \end{cases}
    \label{eq:wind_output}
\end{equation}\\

This gives the power generated by a typical wind turbine in each grid square.

\section{Our method vs.\ renewables.ninja}
\label{appendix:validation}

While reanalysis datasets are routinely used to estimate solar and wind generation in power systems models, there are questions surrounding the extent to which these datasets can be relied upon to accurately predict wind and solar electrical output at a given location. This is the problem discussed by Pfenninger and Staffell in \cite{pfenninger_solar} and \cite{staffell_wind}. Using similar methods to the ones described here in the Methods section, they map weather data onto the locations of existing renewables fleets around Europe to estimate the total renewable output in each European country over time. They then derive national calibration factors by comparing these to the reported national wind and solar outputs in each country.

We note that biases in the underlying weather data are one of a number of potential causes of this discrepancy between estimated national output and reported national output. Other possible contributions to the discrepancy could be imperfect interpolation of MERRA-2 data (which only represents an average over a wide area) onto specific generation sites, imperfect models for calculating wind and solar output from the underlying weather data, $<$100\% availability factors for wind and solar generation facilities in real life, and sampling effects due to the distribution of generation sites.

To address concerns about biases in the underlying weather data affecting the validity of our conclusions, we have repeated our analysis using a dataset made available by Pfenninger and Staffell on the renewables.ninja site \cite{pfenninger_solar, staffell_wind}. The dataset contains calibrated, hourly capacity factors for most European countries over a 32-year period between 1985--2016. This includes every country in our dataset apart from Montenegro and Serbia. We performed a weighted average of these national capacity factors (weighted by national installed capacity in 2022 \cite{country_areas_world_bank}) to arrive at an average European generation figure for each hour.

The major differences between our dataset and the renewables.ninja datset can be summarised as follows:\\

1) The renewables.ninja dataset applies country-wide calibration factors to the underlying MERRA-2 data.\\

2) The renewables.ninja dataset uses the 2016 operating fleet to calculate the capacity factors, rather than assuming that the entire fleet consists of modern wind turbines and solar panels as we have done in the paper.\\

3) When calculating capacity factors for each country, the renewables.ninja dataset uses the locations of the 2016 operating fleet rather than an area-average across the country.\\

4) When calculating European-average capacity factors for the renewables.ninja dataset, we weight each country by 2022 installed capacity rather than by area (as is done in the main text).\\

5) The renewables.ninja dataset runs for 32 years rather than 43 years.\\

We repeated the entire analysis of the paper using the renewables.ninja dataset and generated equivalent plots to the ones shown in the paper using the new dataset. These plots are shown in the Supplementary Information. While neither dataset is perfect, we find that our results are insensitive to the choice of dataset used. This gives us confidence that the results presented here reflect the characteristics of a highly renewable, interconnected European grid, and not the idiosyncrasies of the underlying weather data.

\section{List of Countries Included in the Analysis}
\label{section:countries}

\begin{itemize}
    \item Andorra (microstate)
    \item Austria
    \item Belgium
    \item Bulgaria
    \item Croatia
    \item Czech Republic
    \item Denmark
    \item Estonia
    \item Finland
    \item France
    \item Germany
    \item Greece
    \item Hungary
    \item Italy
    \item Latvia
    \item Liechtenstein (microstate)
    \item Lithuania
    \item Luxembourg
    \item Monaco (microstate)
    \item Montenegro
    \item Netherlands
    \item North Macedonia
    \item Norway
    \item Poland
    \item Portugal
    \item Romania
    \item San Marino (microstate)
    \item Serbia
    \item Slovakia
    \item Slovenia
    \item Spain
    \item Sweden
    \item Switzerland
    \item United Kingdom
    \item Vatican City (microstate)
\end{itemize}
NB: microstates have no published demand data on the ENTSO-E platform, but we assume demand data for these countries is included in the reported numbers for surrounding countries (e.g it is assumed that Monaco demand data is included in reported French data).

\bibliographystyle{elsarticle-num} 
\bibliography{main.bib}

\end{document}